\documentclass[reprint,superscriptaddress,amsmath,amssymb,aps,prl,floatfix]{revtex4-1}
\usepackage{graphicx}
\usepackage{bm}
\usepackage{hyperref}
\usepackage{todonotes}
\usepackage{nicefrac}

\pdfpageattr{/Group <</S /Transparency /I true /CS /DeviceRGB>>}

\begin{document}
\newcommand{\PdCrO}{PdCrO\textsubscript{2}}

\title{Magnetic frustration and spontaneous rotational symmetry breaking in \PdCrO{}}

\author{Dan Sun}
\thanks{These authors contributed equally.}
\affiliation{Max Planck Institute for Chemical Physics of Solids, N\"{o}thnitzer Str 40, 01187 Dresden, Germany}
\affiliation{Los Alamos National Laboratory, Los Alamos, NM 87545, U.S.A.}
\author{Dmitry A. Sokolov}
\thanks{These authors contributed equally.}
\affiliation{Max Planck Institute for Chemical Physics of Solids, N\"{o}thnitzer Str 40, 01187 Dresden, Germany}
\author{Jack Bartlett}
\affiliation{Max Planck Institute for Chemical Physics of Solids, N\"{o}thnitzer Str 40, 01187 Dresden, Germany}
\affiliation{SUPA, School of Physics and Astronomy, University of St. Andrews, St. Andrews KY16 9SS, United Kingdom}
\author{Jhuma Sannigrahi}
\affiliation{Max Planck Institute for Chemical Physics of Solids, N\"{o}thnitzer Str 40, 01187 Dresden, Germany}
\author{Seunghyun Khim}
\affiliation{Max Planck Institute for Chemical Physics of Solids, N\"{o}thnitzer Str 40, 01187 Dresden, Germany}
\author{Pallavi Kushwaha}
\affiliation{Max Planck Institute for Chemical Physics of Solids, N\"{o}thnitzer Str 40, 01187 Dresden, Germany}
\affiliation{CSIR - National Physical Laboratory.  Dr. K. S. Krishnan Marg, New Delhi 110012, India}
\author{Dmitry D. Khalyavin}
\affiliation{ISIS facility, STFC Rutherford Appleton Laboratory, Chilton, Didcot, OX11 0QX, United Kingdom}
\author{Pascal Manuel}
\affiliation{ISIS facility, STFC Rutherford Appleton Laboratory, Chilton, Didcot, OX11 0QX, United Kingdom}
\author{Alexandra S. Gibbs}
\affiliation{ISIS facility, STFC Rutherford Appleton Laboratory, Chilton, Didcot, OX11 0QX, United Kingdom}
\author{Hidenori Takagi}
\affiliation{Max Planck Institute for Solid State Research, Heisenbergstrasse 1, 70569 Stuttgart, Germany}
\affiliation{Institute for Functional Matter and Quantum Technologies, University of Stuttgart, Pfaffenwaldring 57, 70569 Stuttgart, Germany}
\affiliation{Department of Physics, The University of Tokyo, 7-3-1 Hongo, Bunkyo-ku, Tokyo 113-0022, Japan}
\author{Andrew P. Mackenzie}
\affiliation{Max Planck Institute for Chemical Physics of Solids, N\"{o}thnitzer Str 40, 01187 Dresden, Germany}
\affiliation{SUPA, School of Physics and Astronomy, University of St. Andrews, St. Andrews KY16 9SS, United Kingdom}
\author{Clifford W. Hicks}
\affiliation{Max Planck Institute for Chemical Physics of Solids, N\"{o}thnitzer Str 40, 01187 Dresden, Germany}

\date{\today}

\begin{abstract}

In the triangular layered magnet \PdCrO{} the intralayer magnetic interactions are strong, however the lattice structure frustrates interlayer interactions. In spite of this,
long-range, 120$^\circ$ antiferromagnetic order condenses at $T_N = 38$~K. We show here through neutron scattering measurements under in-plane uniaxial stress and in-plane magnetic field that this
occurs through a spontaneous lifting of the three-fold rotational symmetry of the nonmagnetic lattice, which relieves the interlayer frustration. We also show through resistivity measurements that
uniaxial stress can suppress thermal magnetic disorder within the antiferromagnetic phase.

\end{abstract}

\maketitle

At 24 K, solid oxygen undergoes a simultaneous N\'{e}el transition and rhombohedral to monoclinic structural transition~\cite{DeFotis81, Stephens86}. The structural transition is driven by magnetic
frustration: the monoclinic distortion introduces a preferred direction that relieves interlayer frustration~\cite{Rastelli88}. The delafossite compound \PdCrO{} is also a rhombohedral system with
interlayer magnetic frustration. The Cr sites in each layer are triangularly coordinated, and host $S = \frac{3}{2}$ spins that start to arrange themselves into short-range, $120^\circ$
antiferromagnetic order at 200--300~K~\cite{Takatsu09, Hicks15, Le18, Arsenijevic16}. However Cr sites in each layer are centered between the Cr sites in adjacent layers, which frustrates the
interlayer coupling. As the temperature is reduced to just above $T_N = 38$~K, the in-plane correlation length grows to $\sim$20 lattice spacings without appearance of interplane
coherence~\cite{Billington15, Ghannadzadeh17}. Then at $T_N$ the layers lock together to form long-range order at a transition that appears to be weakly first-order~\cite{Hicks15, Takatsu10JPhys}. By
analogy with solid oxygen and a related compound, CuCrO$_2$~\cite{Aktas13}, this could through a spontaneous rotational symmetry breaking and associated structural distortion that relieves interlayer
frustration. However so far no structural distortion has been detected in \PdCrO{}~\cite{Takatsu14, Le18}. 

It is an important point to resolve, to understand the mechanisms by which magnetic order condenses on frustrated lattices, and here we take a different approach: using symmetry-breaking fields to
polarize domains, if they are present.  We employ neutron scattering measurements under in-plane uniaxial stress and magnetic field, and resistivity measurements under uniaxial stress. In \PdCrO{} the
CrO\textsubscript{2} layers are Mott insulating, but are interleaved with highly conducting Pd sheets~\cite{Sunko1809}. Comparison with PdCoO\textsubscript{2}, which is nonmagnetic but otherwise has a
nearly identifal Fermi surface (including the Fermi velocity) to \PdCrO{}, shows that magnetic scattering is the largest component of the inelastic resistivity of \PdCrO{}~\cite{Hicks15}. 

In Fig.~1(a) we illustrate the nonmagnetic lattice. The nonmagnetic unit cell contains three layers: the offset of the Cr layers with respect to each other introduces an ABCABC stacking. The $\sqrt{3}
\times \sqrt{3}$ reconstruction associated with $120^\circ$ magnetic order is observed in quantum oscillation~\cite{Hicks15, Ok13}, angle-resolved photoemission~\cite{Noh14, Sunko1809}, and neutron
data~\cite{Takatsu09, Takatsu14}; early signs of it appear at $\sim$60~K~\cite{Daou15}. The neutron data also suggest a double-$k$ magnetic structure (where $k$ is a propagation vector of the magnetic
structure), with simultaneous ferroic and antiferroic interplane correlations. 

The study of Takatsu \textit{et al} (Ref.~\cite{Takatsu14}) finds three models that give good fits to observed neutron scattering intensities. One of them (model \#2 in that study) is mixture of two
single-$k$ magnetic phases that are not related by symmetry, and has magnetization that varies strongly from site to site. Varying magnetization is not compatible with the strong on-site Hund's-rule
polarisation that drives the Mott insulating behavior, so this model is unlikely to be realized. The other two models (\#3 and \#4) incorporate the two $k$'s through alternating vector chirality. That
is, the direction of rotation of the spins on moving from site to site alternates from layer to layer. In both, the three-fold rotational symmetry of the lattice is broken, implying an associated
monoclinic or triclinic lattice distortion and the presence of domains. Domains are a complication in analysis of magnetic structures, and in Ref.~\cite{Takatsu14} it was assumed for analysis that
each of the three domain types were equally populated. Models \#3 and \#4 are closely related, differing by modest collective spin rotations, and we illustrate in Fig.~1(b) a magnetic structure that
is a simplified version of both. It lifts rotational symmetry in the same way and gives only a marginally worse match to the reflection intensities reported in Ref.~\cite{Takatsu14} (we quantify this
statement later), so for discussion we refer to this structure for now and explain possible refinements later. The rotational symmetry breaking appears in two aspects of the structure: the spins lie
in the $yz$ plane, and from plane to plane the magnetic order shifts along the $y$ axis, as indicated by the blue dashed line in the figure.

\begin{figure}[ptb]
\includegraphics[width=85mm]{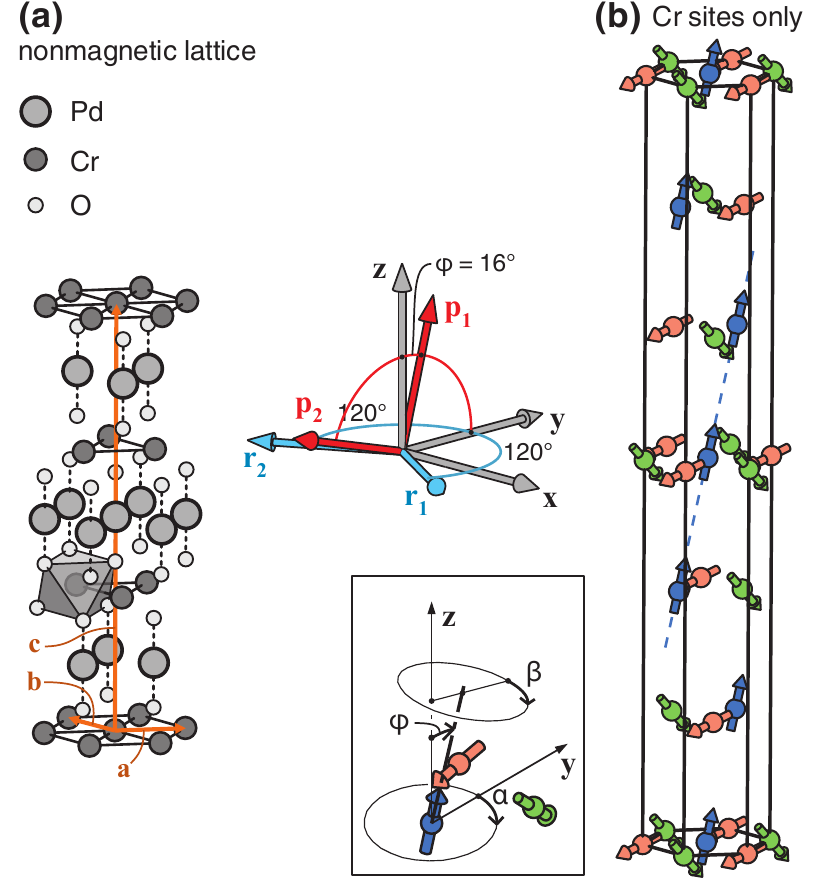}
\caption{\label{magStructure}\textbf{(a)} The delafossite structure. Here, $\mathbf{\hat{x}}$ is along a $\langle 100 \rangle$ direction (a Cr-Cr bond direction), and
$\mathbf{\hat{z}}$ along the $c$ axis. In the neutron measurements, field or pressure was applied along the $y$ axis, and the scattering plane was the $xz$ plane.  \textbf{(b)} A magnetic structure
that gives a good fit to the 58 scattering intensities reported in Ref.~\cite{Takatsu14}. The spins lie in the $yz$ plane, and are colored by orientation. Alternating vector chirality is indicated by
reversal of the positions of the green and pink spins from layer to layer.  Collective spin rotations can be parametrized as shown in the box: $\phi$ and $\alpha$ are the polar and azimuthal angles of
the blue spin, and $\beta$ the angle of rotation of the spin plane about the axis of the blue spin, which sets the orientations of the other two spins. Here, $\phi = 16^\circ$, and $\alpha = \beta =
0$.}
\end{figure}

\begin{figure}[ptb]
\includegraphics[width=85mm]{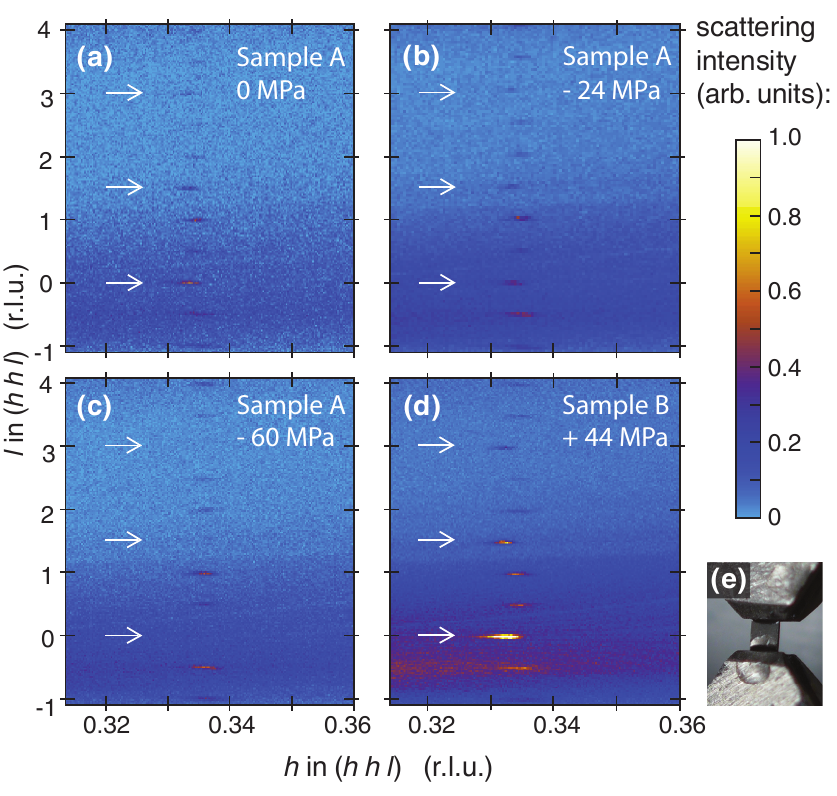}
\caption{\label{neutronScattering}Magnetic reflections at 2~K at various uniaxial stresses $\sigma$ applied along a $\langle 1 \bar{1} 0 \rangle$ direction, \textit{i.e.} the $y$ axis in Fig.~1.
Negative values indicate compression. The $l=0$, $\frac{3}{2}$, and $3$ reflections are indicated by white arrows. Under compression these reflections fade together, and under tension they are
strengthened together. This indicates that these peaks originate from a single domain type. \textbf{(e)} Photograph of sample A; this sample is 0.97~mm wide and 0.11~mm thick.}
\end{figure}

All data reported here are on crystals grown by the NaCl flux method in a sealed quartz tube, as reported in~\cite{Takatsu10JCG}. Three samples, labelled A, B, and C, were studied with neutrons.
Samples A and B were cut and polished to respective dimensions $1.65 \times 0.97 \times 0.11$~mm$^3$ and $2.13 \times 1.30 \times 0.07$~mm$^3$, and mounted into holders that provide the necessary
mechanical protection to apply in-plane stresses to plate-like samples. Force was applied using a mechanical spring which was adjusted at room temperature, so samples were cooled under nearly constant
stress. The cooling rate was rapid: $\sim$20~K/min.  A photograph of sample A is shown in Fig.~\ref{neutronScattering}(e). Most of the sample is exposed and experiences the full applied force, however
the neutrons do also penetrate into the ends of the sample, which are embedded in epoxy and are under lower stress than the central portion. Sample A was probed under compressive stresses of up to
$\sigma = -108$~MPa, and sample B a tensile stress of +44~MPa. (We use negative/positive values to indicate compression/tension.) Force was applied along a $\langle 1 \bar{1} 0 \rangle$ direction,
corresponding to the $y$ axis in Fig.~1, and the scattering plane was the $xz$ plane. Sample C was studied under magnetic field applied along a $\langle 1 \bar{1} 0 \rangle$ direction. All neutron
measurements were performed using the WISH diffractometer at the ISIS spallation neutron source. 

Results from samples A and B are shown in Fig.~\ref{neutronScattering}. At zero stress, scattering peaks appear at $(hkl) = (\frac{1}{3} \frac{1}{3} l)$ r.l.u. for every half-integer $l$. The
reflections are referenced to the three-layer nonmagnetic unit cell indicated in Fig.~1(a), so a two-layer periodicity, for example, yields a reflection at $l = \frac{3}{2}$. The reflections at $l =
0$, $\frac{3}{2}$, and 3 faded as compressive stress was applied. They were weaker at -24~MPa, and disappeared almost entirely by -60~MPa. Conversely, tensile stress strengthened these peaks and
suppressed the others. This rapid evolution with stress almost certainly indicates polarization of a domain structure. The applied strains are tiny: for a typical Young's modulus for an oxide material
of $\sim$150~GPa~\cite{Paglione02, Roa07}, stress $\sigma = 60$~MPa corresponds to strain $\varepsilon \sim 4 \cdot 10^{-4}$. 

The intensities of reflections spaced by $\Delta l = \frac{3}{2}$ evolve together with applied stress, indicating three domain types that give, for $(h k)$ = $(\frac{1}{3} \frac{1}{3})$, reflections
at $l = \{0, \frac{3}{2}, 3, ...\}$, $\{\frac{1}{2}, 2, \frac{7}{2}, ...\}$, and $\{1, \frac{5}{2}, 4, ...\}$, respectively. This is as expected for the magnetic structure of Fig.~1(b). The
alternating vector chirality means the spin components parallel to the vector $\mathbf{p_1}$ [see Fig.~1(b)] are ferroically aligned between the layers, yielding the reflection at $(\frac{1}{3}
\frac{1}{3} 0)$, while components along $\mathbf{p_2} \perp \mathbf{p_1}$ are antiferroically aligned, yielding the reflection at $(\frac{1}{3} \frac{1}{3} \frac{3}{2})$. $(\frac{1}{3} \frac{1}{3} 0)$
lies along the $+x$ direction in the figure, while in the illustrated magnetic structure the magnetic order shifts along the $y$ direction from layer to layer. Because these directions are
perpendicular there is no offset of the reflections along $l$, and this structure yields reflections at $\{0, \frac{3}{2}, 3, ...\}$.  However in the absence of an applied symmetry-breaking field the
interplane ordering vector could equivalently lie in the $\mathbf{r_1}$-$\mathbf{z}$ or $\mathbf{r_2}$-$\mathbf{z}$ planes, where $\mathbf{r_1}$ and $\mathbf{r_2}$ are vectors rotated from
$\mathbf{y}$ by $120^\circ$ [again, see Fig.~1(b)]. Because $\mathbf{r_1}$ and $\mathbf{r_2}$ are not perpendicular to $(\frac{1}{3} \frac{1}{3} 0)$, an offset is introduced: $r_1z$ domains yield
reflections at $l = \{\frac{1}{2}, 2, \frac{7}{2}, ...\}$ and $r_2z$ domains at $\{1, \frac{5}{2}, 4, ...\}$.  $r_1z$ and $r_2z$ domains are symmetrically equivalent under $y$-axis uniaxial stress,
and so would be suppressed or favored together, as observed. We note that similar analysis could be done for $(h k) = (\frac{1}{3}\,-\!\frac{2}{3})$ and $(-\frac{2}{3} \frac{1}{3})$, however these reflections
were not accessible in this measurement. 

Data under magnetic field are shown in Figs.~3(a-b). These data were collected at $T = 1.5$~K, without thermal cycling between the different fields. A 13~T field applied along the $y$ axis suppresses
the reflection at $(\frac{1}{3} \frac{1}{3} 0)$ completely, while the reflection at $(\frac{1}{3} \frac{1}{3} 1)$ remains. Evidently, applied field along $\mathbf{y}$ favors, like
compression along $\mathbf{y}$, the $r_1z$ and $r_2z$ domains. Resistivity data presented below indicate considerable hysteresis in domain re-orientation at low temperatures, so if the sample
were field-cooled the domains would likely polarize under considerably smaller fields.

\begin{figure}[ptb]
\includegraphics[width=85mm]{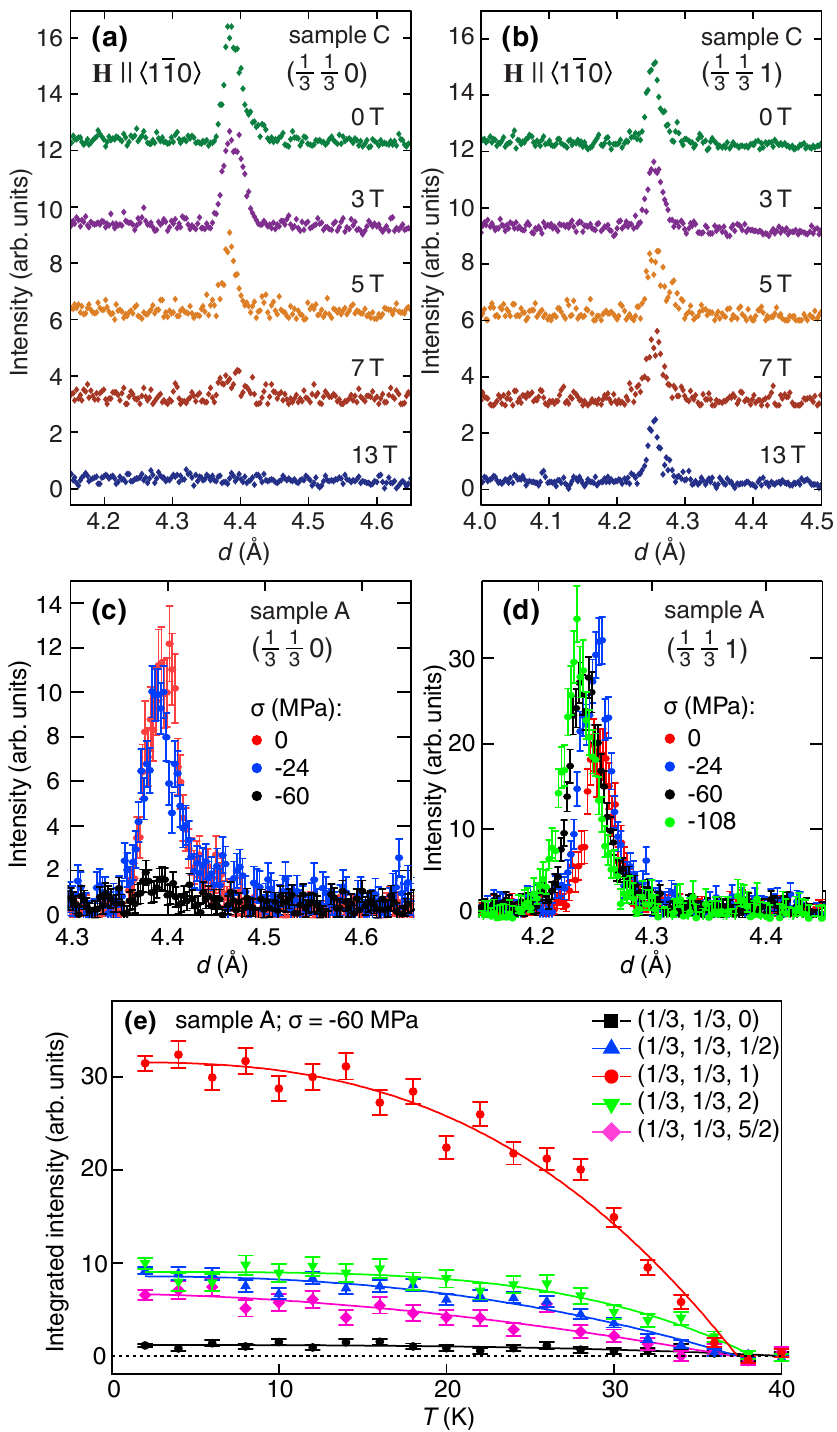}
\caption{\label{neutronTdependence}\textbf{(a-b)} Field dependence of the scattering intensities at $(\frac{1}{3} \frac{1}{3} 0)$ and $(\frac{1}{3} \frac{1}{3} 1)$ for sample C, at 1.5~K.
Field is applied along $(1\bar{1}0)$, \textit{i.e.} the $y$ axis in Fig.~1. $d$ is the real-space periodicity. \textbf{(c-d)} Stress dependence of the magnetic scattering intensities at $(\frac{1}{3}
\frac{1}{3} 0)$ and $(\frac{1}{3} \frac{1}{3} 1)$ of sample A, at 2~K. \textbf{(e)} Temperature dependence of the integrated intensities of various reflections under $\sigma = -60$~MPa. A
temperature-independent background is subtracted from each scan.}
\end{figure}

Fig.~\ref{neutronTdependence}(e) shows integrated scattering intensities as a function of temperature for $\sigma = -60$~MPa. The $l=0$ peak was the most intense at $\sigma=0$, while under
$\sigma=-60$~MPa it remains suppressed up to $T_N$. We conclude that the magnetic structure remains polarized up to $T_N$, without thermal excitation of disfavored domains.

\begin{figure}[ptb]
\includegraphics[width=85mm]{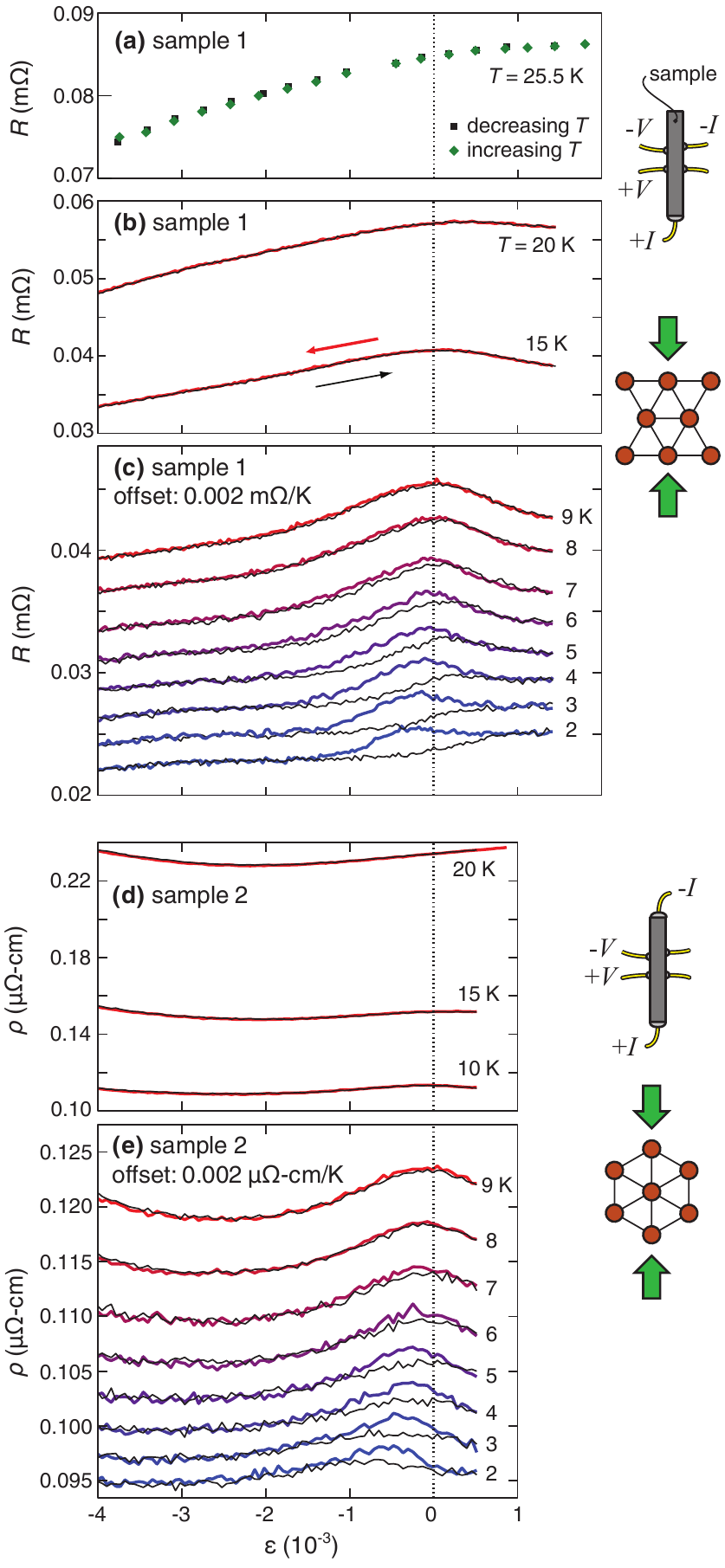}
\caption{\label{resistivity}Resistance versus strain of two samples of \PdCrO{}. \textbf{(a-c)} A sample cut along a $\langle 1 \bar{1} 0 \rangle$ direction. One of the contacts broke during cooling
-- the sample was under 1~mm long -- so we used the contact configuration indicated in the figure. Data in panel (a) are from temperature ramps at fixed strain from above $T_N$, and in panels (b) and
(c) from strain ramps at fixed temperature. \textbf{(d-e)} A sample cut along a $\langle 100 \rangle$ direction. For all panels, $\varepsilon=0$ is taken as the location of the peak in $\rho$ at 9~K.
Colored lines are decreasing-strain, black lines are increasing-strain ramps. The ramp rate for all curves is below $10^{-6}$~s$^{-1}$.}
\end{figure}

Therefore, if resistivity is measured while ramping the applied stress across zero, it is reasonable to hypothesize that a step-like feature should appear when the domains re-orient, corresponding to
the resistive anisotropy within a domain. Detectable resistivity anisotropy is more likely to appear in the inelastic than the elastic component, because ARPES data indicate that the Fermi surface
remains highly symmetric below $T_N$~\cite{Noh14, Sobota13}. (Although the samples will not have been detwinned in the ARPES studies, the observed Fermi surfaces remain sharp and highly isotropic
modulo six-fold rotation symmetry.) To measure $\rho$($\varepsilon$, $T$) we prepared the samples as beams and mounted them into piezoelectric-based uniaxial stress cells, as has been described
previously~\cite{Hicks14, Barber18, Barber19}. A technical difference between the neutron and resistivity data is that in the former the controlled variable is stress, while in the latter it is
strain. This is because for the neutron measurements force was applied using springs with spring constants much lower than those of the samples, while for the resistivity measurements piezoelectric
actuators with a very high combined spring constant were used. The proportionality constant between uniaxial stress and strain is the Young's modulus. We report data from three samples, two oriented
along a $\langle 1 \bar{1} 0 \rangle$ direction, \textit{i.e.} bisecting the Cr-Cr bond axes, and one along a $\langle 100 \rangle$ direction, \textit{i.e.} along a Cr-Cr bond direction. Results from
two samples are shown in Fig.~\ref{resistivity}. 

For both orientations, $\rho(\varepsilon)$ at 9~K has a relatively sharp peak, and a capacitive displacement sensor built into the apparatus indicates that at the peak $|\varepsilon| < 10^{-3}$. We
therefore fix $\varepsilon=0$ as the location of the peak in $\rho(\varepsilon)$ at 9~K. This assignment is further supported by the appearance of a hysteresis loop at lower temperatures centered on
this strain; larger hysteresis in domain reorientation is expected at low temperatures. Because the intrinsic resistivity at low temperature is probably highly isotropic, the changes in resistivity at
these temperatures are most likely due to changes in magnetic disorder driven by domain reversal.

No step-like feature in $\rho(\varepsilon)$ is resolvable above $\sim$9~K for either stress orientation, despite the clear indication from neutron data of polarizable domains up to
$T_N$. In principle, it is possible that domains are strongly pinned, and do not re-orient in strain ramps. Therefore we also performed temperature ramps from above $T_N$ down to 25.5~K, conditions
under which the neutron data indicate unambiguously that the magnetic order polarizes under strains well below $10^{-3}$ for any plausible assumption about the Young's modulus of \PdCrO{}. Again,
there is no resolvable step at or near $\varepsilon=0$; see Fig.~\ref{resistivity}(a). We conclude that any intrinsic resistive anisotropy within a domain is below our resolution,
consistent with the symmetry breaking being magnetically- rather than electronically-driven.

Strain does however have a strong effect on $\rho$: the presence of the peak at $\varepsilon=0$, especially prominent for $\sim 4 < T < 15$~K, indicates that there is magnetic
disorder that can be suppressed by uniaxial stress. The peak broadens as $T$ is raised, indicating that it is thermal disorder. The sharpness of the peak at lower
temperatures is striking. Even though the magnetic structure lifts the triangular symmetry of the lattice, it appears that being close to triangular symmetry gives a high susceptibility to thermal
disorder, through the existence of one or more low-energy spin wave modes.

To discuss which degrees of freedom yield disorder that might be suppressed by uniaxial stress, we parametrize the spin rotation angles as shown in Fig.~1(b). Following previous work~\cite{Takatsu14,
Le18}, the order within each layer is taken to be co-planar $120^\circ$ order, favored by strong intralayer interactions, and $\phi_i$ and $\alpha_i$ are the polar and azimuthal angles, respectively,
of a selected spin within layer $i$. There is formally a third degree of freedom, $\beta_i$, the azimuthal angle of the spin plane about this reference spin, which sets the orientations of the other
two spins, however $\alpha$ and $\beta$ become indistinguishable parameters in the limit $\phi \rightarrow 0$. We calculate a goodness-of-fit $S' = [\chi^2/(n-1)]^{1/2}$ to the scattering intensities
reported in Ref.~\cite{Takatsu14}, where $n = 58$ is the number of reported intensites and 1 is subtracted for an overall scaling factor to compare measured and calculated intensities, and as in that
reference we average over the three domain types. $S'$ below $\sim$1.3 indicates a good fit. The structure shown in Fig.~1(b), which has $\phi = 16^\circ$ and $\alpha = \beta = 0$ in each layer, gives
$S'=1.06$. The refinement in model \#3 of Ref.~\cite{Takatsu14} is layer-to-layer variation in $\phi$, and in model \#4 in $\alpha$; these refinements give $S' = 1.00$ and $0.99$, respectively.

$\phi = 16^\circ$ is small enough that the distinction between $\alpha$ and $\beta$ is not very meaningful, and so we now fix $\beta = 0$ and allow rotations of the spins out of the $yz$ plane only
through nonzero $\alpha$. We consider nonzero $\beta$ in the appendix. In Ref.~\cite{Le18}, a spin wave gap of 0.4~meV was observed in inelastic neutron scattering, and reproduced in calculations of
classical dipole-dipole interactions for spin waves in $\alpha$. 0.4~meV corresponds to a temperature of $\sim$4~K, which is approximately the temperature at which a peak in $\rho(\varepsilon)$
becomes discernible. Therefore it appears likely that the effect of uniaxial stress is to increase the spin wave gap for spin excitations out of the $yz$ plane. 

To probe how far spins might fluctuate away from the $yz$ plane, we calculate intensities from a 300-layer magnetic cell within which $\alpha$ is chosen randomly in each layer, from a Gaussian
distribution centered on $\alpha = 0$ and with standard deviation $\sigma_\alpha$.  At 2~K, the best match to observed intensities is obtained when $\sigma_\alpha = 10$--$15^\circ$. However the
improvement on locking the spins into the $yz$ plane is marginal: $S'$ decreases from 1.06 to 0.97. At 30~K however the best match is obtained with $\sigma_\alpha \approx 20^\circ$, and now $S'$
decreases from 1.30 to 1.10.  Therefore the neutron data of Ref.~\cite{Takatsu14} suggest with modest statistical confidence that in unstressed \PdCrO{} the spin wave modes allowing rotation of the
spins out of the $yz$ plane are softer than other modes.  We note that a calculation with $\alpha$ alternating regularly from layer to layer gives statistically indistinguishable results, however a
hypothesis of random variation is more consistent with thermal disorder.

In conclusion, we have shown through neutron scattering measurements under applied uniaxial stress and applied magnetic field that the magnetic order of \PdCrO{} relieves interlayer frustration by
spontaneously lifting the three-fold rotational symmetry of the nonmagnetic lattice. This rotational symmetry breaking is not detectable in resistivity, showing that it is a magnetic rather than an
electronic instability. Resistivity measurements indicate the presence of low-energy spin wave modes when the lattice is close to triangularly symmetric, that yield fluctuations in the magnetic order
that are suppressed through in-plane uniaxial stress. More generally, the ability to polarize magnetic domains through uniaxial stress will in the future allow greater precision in the determination
of magnetic structures, by eliminating domain population as a degree of freedom.

We acknowledge helpful discussions with Erez Berg, Takashi Oka, and Onur Erten. We acknowledge the financial support of the Max Planck Society. Experiments at the ISIS Pulsed Neutron and Muon Source
were supported by a beamtime allocation from the Science and Technology Facilities Council under RB1520411, DOI 10.5286/ISIS.E.67774469 for the field work, and RB1800029, DOI 10.5286/ISIS.E.90605228
for the stress work. Raw data are available at \textit{to be determined}.
\\
\\
\noindent \textbf{Appendix.}

\begin{figure}[ptb]
\includegraphics[width=85mm]{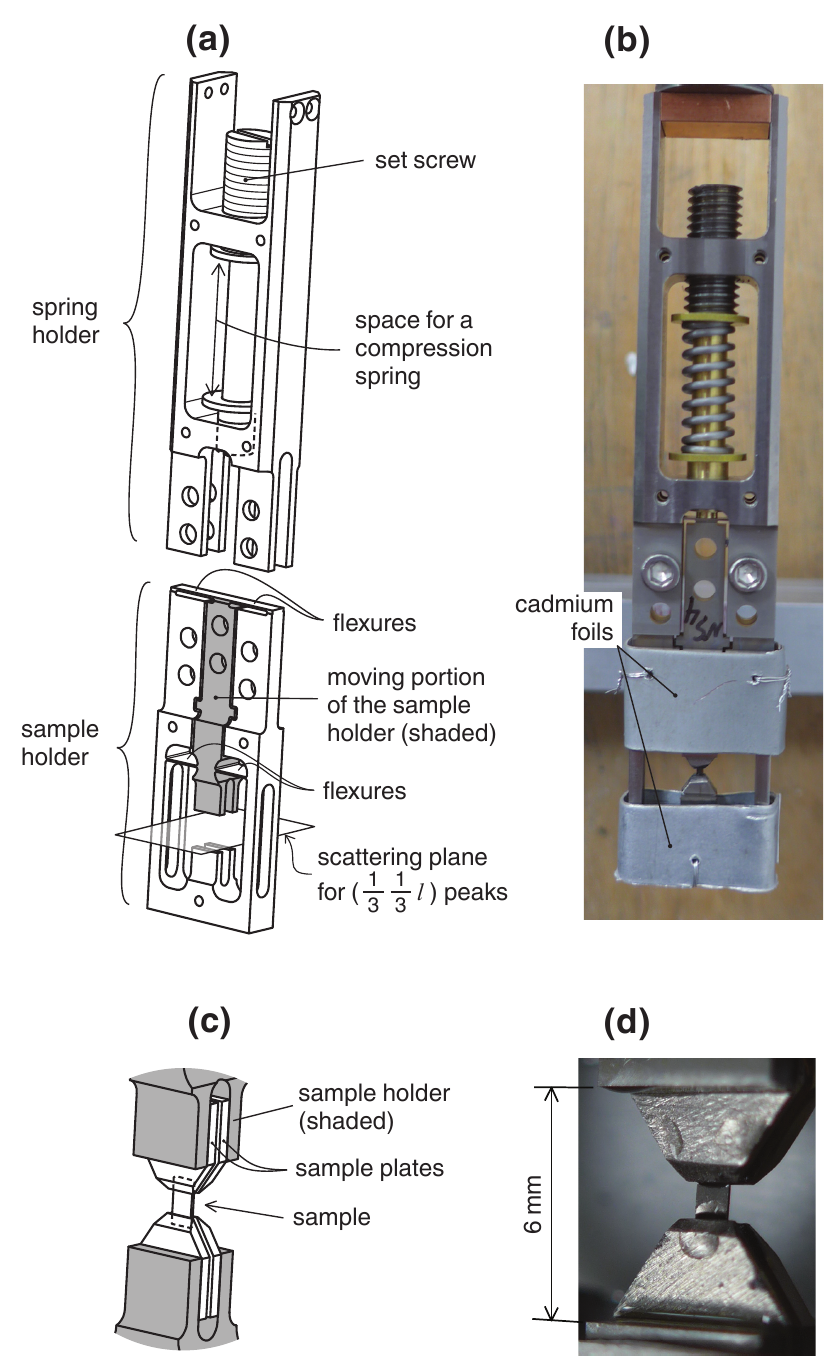}
\caption{\label{exptSetup_neutrons} The experiment setup for neutron scattering. \textbf{(a)} Drawing of the spring and sample holders. The configuration of the spring holder in this drawing is for
applying compressive load onto the sample, however a tension spring can also be installed. Force is applied to a moving portion of the sample holder, whose motion is constrained by flexures to be
longitudinal with respect to the sample. This protects the sample from inadvertent transverse or twisting forces. \textbf{(b)} Photograph of this setup, including cadmium foils used to absorb stray
neutrons. \textbf{(c)} Drawing of a mounted sample. The sample is plate-like, and each end is secured with epoxy between two sample plates. The end portions of the sample, embedded in the epoxy, will
be under lower stress than the central, exposed portion. \textbf{(d)} A photograph of sample A, taken shortly after removal from the cryostat.} \end{figure}

In this appendix, we provide more details of the experiment setup, details of the calculation of magnetic neutron reflection intensites, and some supplementary data.  Fig.~\ref{exptSetup_neutrons}
illustrates the uniaxial stress apparatus that we used here for neutron scattering. A plate-like sample geometry is compatible with application of very high uniaxial stress, with high stress
homogeneity. In our apparatus, samples are held in detachable holders (allowing rapid sample exchange during a beamtime) that leave as much space around the sample exposed as possible. The holder
incorporates flexures that protect the sample from inadvertent twisting or transverse forces; this is essential because the samples are thin and mechanically fragile. The holder slots into a spring
holder, which holds either a compression or tension spring to apply force to the sample. A set screw is used to adjust the force; the force applied was determined by multiplying the spring constant of
the spring, supplied by the manufacturer, with the applied displacement, which was measured with a ruler. As the set screw can only be adjusted at room temperature, samples are cooled under
approximately constant stress. The spring constant of the springs used will have increased by $\sim$10\% with cooling to cryogenic temperatures, but that is not essential to the work presented here.

The magnetic reflections are indexed to a 3-site nonmagnetic unit cell. The lattice vectors of this cell are
\begin{eqnarray*}
\mathbf{a} & = & (a,0,0), \\
\mathbf{b} & = & (-\frac{a}{2}, \frac{a\sqrt{3}}{2}, 0), \\
\mathbf{c} & = & (0, 0, c).
\end{eqnarray*}
$a = 2.93$~\AA{} is the Cr-Cr interatomic spacing, and $c = 18.087$~\AA{} spans three layers. The reciprocal lattice vectors of this cell are
\begin{eqnarray*}
\mathbf{h_3} & = & 2 \pi (\frac{1}{a}, \frac{1}{a \sqrt{3}}, 0), \\
\mathbf{k_3} & = & 2 \pi (0, \frac{2}{a \sqrt{3}}, 0), \\
\mathbf{l_3} & = & 2 \pi (0, 0, \frac{1}{c}).
\end{eqnarray*}

The atomic positions in the first layer are given by
\begin{eqnarray*}
\mathbf{x}_1 & = & (0, 0, 0), \\
\mathbf{x}_2 & = & (\frac{a}{2}, \frac{a \sqrt{3}}{2}, 0), \\
\mathbf{x}_3 & = & (-\frac{a}{2}, \frac{a \sqrt{3}}{2}, 0). \\
\end{eqnarray*}
The magnetic moments on these sites are given by
\begin{eqnarray*}
\mathbf{M}_1 & = & (\sin(\phi_1)\sin(\alpha_1), \sin(\phi_1)\cos(\alpha_1), \cos(\phi_1)), \\
\mathbf{M}_2 & = & \cos(\frac{2\pi}{3})\mathbf{M}_1 + \sin(\frac{2\pi}{3})\mathbf{M}_1 \times \mathbf{\hat{r}}, \\
\mathbf{M}_3 & = & \cos(-\frac{2\pi}{3})\mathbf{M}_1 + \sin(-\frac{2\pi}{3})\mathbf{M}_1 \times \mathbf{\hat{r}},
\end{eqnarray*}
where
\begin{eqnarray*}
\mathbf{\hat{r}} & = & \Big(\cos(\beta_1)\cos(\alpha_1) - \sin(\beta_1) \sin(\alpha_1) \cos(\phi_1), \\
&& -\cos(\beta_1) \sin(\alpha_1) - \sin(\beta_1) \cos(\alpha_1) \cos(\phi_1), \\
&& \sin(\beta_1) \sin(\phi_1) \Big).
\end{eqnarray*}
The spin rotation angles $\phi$, $\alpha$, and $\beta$ are illustrated in Fig.~1; $\phi$ and $\alpha$ are respectively the polar and azimuthal angle of a reference spin in each layer, and $\beta$ the
azimuthal angle of the spin plane about this reference spin. In the above expressions, the subscript on $\phi$, $\alpha$, and $\beta$ refers to the layer number. $\alpha$ is defined to be zero when
the reference spin lies in the $yz$ plane, and $\beta$ to be zero when the spin plane contains the $z$ axis.

The atomic positions and magnetic moment orientations in subsequent layers are taken as illustrated in Fig.~1. To calculate scattering intensites, we sum over the 18 sites of the magnetic unit cell:
\[
\mathbf{M}(\mathbf{q}) = \sum_{i=1}^{18}e^{i\mathbf{q}\cdot\mathbf{R}_i}\mathbf{M}_i.
\]
Neglecting pre-factors, the scattering intensities are
\[
I(\mathbf{q}) = f(|\mathbf{q}|) \times |\mathbf{M}_\perp(\mathbf{q})|^2,
\]
where f($\mathbf{q})$ is the magnetic form factor of Cr$^{3+}$, and $\mathbf{M}_\perp(\mathbf{q}) = \mathbf{M}(\mathbf{q}) - \mathbf{\hat{q}}(\mathbf{\hat{q}} \cdot \mathbf{M})$. The goodness-of-fit
$S'$ is given by
\[
S' = \left( \frac{1}{57} \sum_{i=1}^{58} \left( \frac{S_{\mathrm{obs}, i} - x S_{\mathrm{calc}, i}}{\sigma_i} \right)^2 \right)^{1/2},
\]
where $x = \sum S_\mathrm{obs} / \sum S_\mathrm{calc}$, and 58 is the number of intensities reported in Ref.~[11]. The magnetic structure, as explained in the main text, lifts the three-fold
rotational symmetry of the nonmagnetic lattice, and so to compare with the intensities reported in Ref.~[11] we average over the three possible rotations of this structure about $\mathbf{\hat{z}}$.

In Fig.~6 we show $S'$ for a fully-specified magnetic structure in which $\phi$, $\alpha$, and $\beta$ are the same in each layer. $\phi$ is seen to be tightly constrained: deviation
from $16^\circ$ by more than $3^\circ$ raises $S'$ above 1.3. $\alpha$ and $\beta$ become indistinguishable parameters in the limit $\phi \rightarrow 0$, and $\phi = 16^\circ$ is small enough that, as
illustrated in the second panel, $\alpha$ and $\beta$ are not tightly constrained mathematically. However the sum $\alpha + \beta$ is tightly constrained around zero, indicating that the best fit is
obtained when the spins lie within the $yz$ plane.

In Fig.~7 we illustrate $S'$ for a model where layer-to-layer fluctuations of the spins out of the $yz$ plane are allowed. This is the model described in the main text: a 300-layer magnetic unit cell
is taken, $\phi$ is fixed at $16^\circ$ in all layers, and $\alpha$ and $\beta$ in each layer are drawn randomly from Gaussian probability distributions centred on $0^\circ$ with standard deviations
$\sigma_\alpha$ and $\sigma_\beta$. Because $\alpha$ and $\beta$ are not highly orthogonal for small $\phi$, $S'$ is found to be nearly constant along lines of constant $\sqrt{\sigma_\alpha^2 +
\sigma_\beta^2}$. As described in the main text, at 2~K the best fit is obtained when the spins fluctuate by $\sim$12$^\circ$ out of the $yz$ plane, but the improvement on locking them into the $yz$
plane ($\alpha = \beta = 0$) is not large: $S'$ decreases from 1.06 to about 0.97. At 30~K the best fit is obtained with $\sqrt{\sigma_\alpha^2 + \sigma_\beta^2} \sim 20^\circ$, and now the improvement
on $\alpha = \beta = 0$ is larger: $S'$ decreases from 1.30 to 1.11. In other words, as temperature is raised the magnetic structure appears to soften through fluctuations of the spins out of the $yz$
plane faster than through other modes.

The integrated intensities of the $(\frac{1}{3}, \frac{1}{3}, l)$ magnetic reflections from sample A at 2 K are given in Table~1.

\begin{table}[h!]
\centering
\begin{tabular}{|c c c c c|}
 \hline
 (h,k,l)  & 0 (MPa) & -24 (MPa) & -60 (MPa) & -108 (MPa)\\ [0.5ex]

 \hline
 \hline
($\frac{1}{3},\frac{1}{3}$,-1) & 14.2(5) & 19.9(4) & 23.5(4) & 28.5(4)\\
($\frac{1}{3},\frac{1}{3}$,-0.5) & 12.0(4) & 15.2(3) & 16.0(3) & 15.3(3)\\
($\frac{1}{3},\frac{1}{3}$,0) & 19.2(7) & 13.7(4) & 3.9(2) & 4.0(2)\\
($\frac{1}{3},\frac{1}{3}$,0.5) & 7.6(5) & 9.8(4) & 12.4(4) & 14.2(4)\\
($\frac{1}{3},\frac{1}{3}$,1) & 26(1) & 32.5(8) & 33.7(8) & 31.0(7)\\
($\frac{1}{3},\frac{1}{3}$,1.5) & 16(1) & 11.5(6) & 3.4(3) & 4.1(3)\\
($\frac{1}{3},\frac{1}{3}$,2) & 8.4(8) & 8.2(5) & 11.8(6) & 18.9(6)\\
($\frac{1}{3},\frac{1}{3}$,2.5) & 6.6(7) & 8.5(5) & 9.1(5) & 2.1(2)\\
($\frac{1}{3},\frac{1}{3}$,3) & 9.4(8) & 7.9(5) & 2.9(3) & 1(1)\\
($\frac{1}{3},\frac{1}{3}$,3.5) & 6.1(6) & 9.0(5) & 6.9(4) & 12.8(5)\\
($\frac{1}{3},\frac{1}{3}$,4) & 9.2(7) & 10.9(5) & 11.9(5) & 13.6(5)\\
[1ex]
 \hline
\end{tabular}
\caption{Integrated intensities of ($\frac{1}{3},\frac{1}{3}$,l) magnetic reflections measured at 2 K under various compressive stresses. Because the temperature-independent background was measured
only at -60~MPa, here we do not subtract off any background. The intensities are normalized to the neutron flux.}
\end{table}

\begin{figure}[ptb]
\includegraphics[width=85mm]{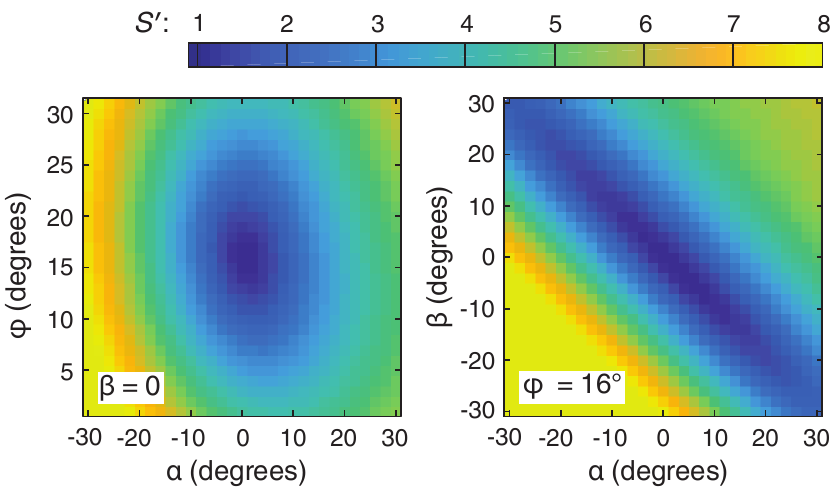}
\caption{\label{SimFig1} Goodness-of-fit $S'$ to the scattering intensities at 2 K reported in Ref.~[11], for a magnetic structure with identical $\phi$, $\alpha$, and $\beta$ in each
layer. In the left-hand panel, $\beta$ is set to zero, and in the right-hand panel $\phi$ is set to $16^\circ$.}
\end{figure}

\begin{figure}[ptb]
\includegraphics[width=85mm]{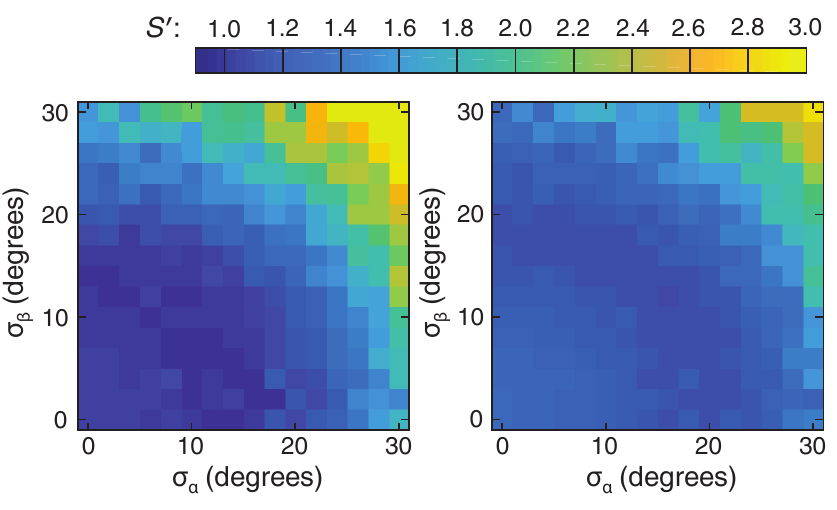}
\caption{\label{simulation2} Goodness-of-fit $S'$ to the scattering intensities at 2~K (left) and 30~K (right) reported in Ref.~[11], taking $\phi = 16^\circ$ in each layer, and $\alpha$ and $\beta$
to vary randomly from layer to layer, drawn from Gaussian distributions centered on $0^\circ$ and with standard deviations $\sigma_\alpha$ and $\sigma_\beta$.}
\end{figure}

\begin{figure}[ptb]
\includegraphics[width=85mm]{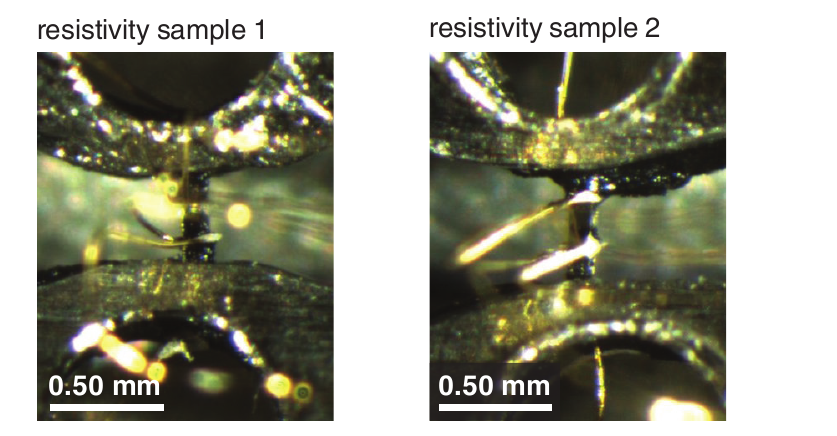}
\caption{\label{resistivitySamplePhotos}Photographs of the two resistivity samples reported in Fig.~4; the left-hand photograph is of the $\langle 1 \bar{1} 0 \rangle$ sample, and the
right-hand of the $\langle 1 0 0 \rangle$ sample.}
\end{figure}

Photographs of the two resistivity samples reported in Fig.~4 are shown in Fig.~\ref{resistivitySamplePhotos}.

\begin{figure}[ptb]
\includegraphics[width=85mm]{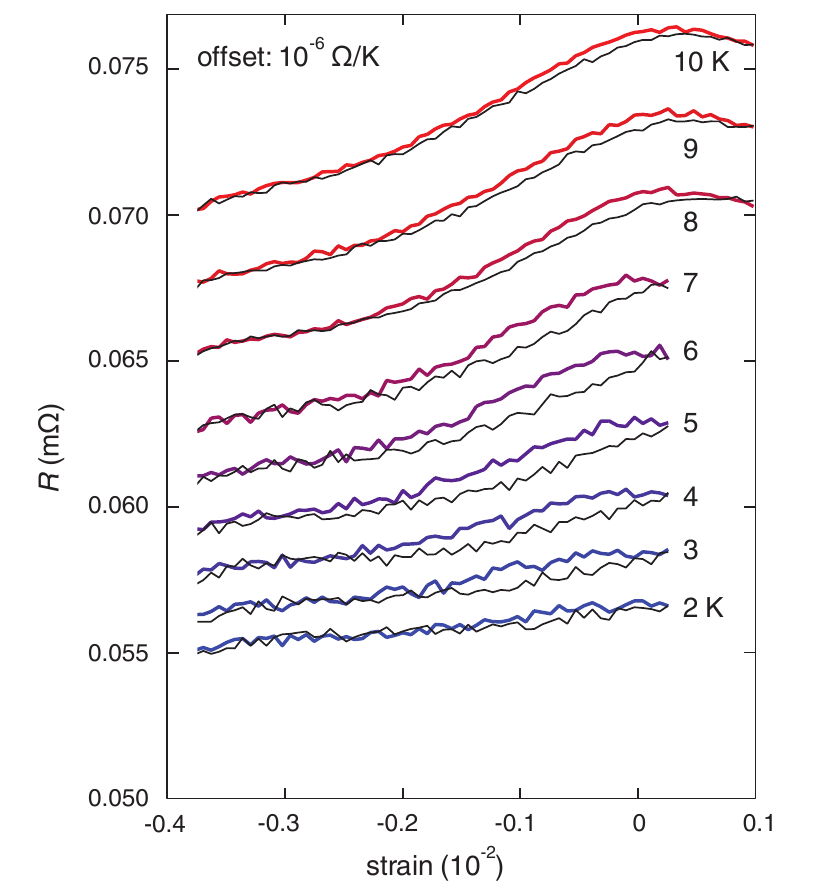}
\caption{\label{resistivitySample3}Resistivity $\rho$ versus strain for a third resistivity sample. Pressure is applied along a $\langle 1 \bar{1} 0 \rangle$ direction.}
\end{figure}

Fig.~\ref{resistivitySample3} shows resistivity versus strain data for a third resistivity sample. Pressure was applied along a $\langle 1 \bar{1} 0 \rangle$ direction, the same as for sample 1 in
Fig.~4. In this sample, there is discernable hysteresis up to 10~K. The hysteresis gradually flattens as $T$ is raised, but at all temperatures extends out to a a rather large strain
of almost $3 \cdot 10^{-3}$.

\begin{thebibliography}{99}
\bibitem{UAHEfootnote}In Ref.~\cite{Takatsu10PRL}, an anomalous Hall effect was reported in \PdCrO{} in the
absence of resolvable magnetization, which was interpreted as indicating nonzero scalar spin chirality, in
other words non-coplanar spins. This anomalous Hall effect however is induced by a nonzero field, so we think
that the possibility that the spins are coplanar in zero field should not be excluded.
\end{thebibliography}
\end{document}